\documentclass{PoS}

\usepackage{bm}        
\usepackage{amssymb}   
\usepackage{amsmath}   
\usepackage{tabularx}   

\newcommand{\beq}{\begin{equation}}
\newcommand{\eeq}{\end{equation}}
\newcommand{\bea}{\begin{align}}
\newcommand{\eea}{\end{align}}
\newcommand{\beas}{\begin{align*}}
\newcommand{\eeas}{\end{align*}}

\newcommand{\Tint}[1]{{\hbox{$\sum$}\!\!\!\!\!\!\!\int\,}_{\!\!\!\!\raise-0.9ex\hbox{$\scriptstyle{#1}$}}}

\title{The in-medium heavy quark potential from quenched and dynamical
   lattice QCD}

\ShortTitle{The in-medium heavy quark potential from quenched and dynamical
   lattice QCD}

\author{Yannis Burnier \\
        Institut de Th\'eorie des Ph\'enom\`enes Physiques, Ecole Polytechnique F\'ed\'erale de Lausanne, CH-1015, Lausanne, Switzerland,\\
        E-mail: \email{yannis.burnier@epfl.ch}}
  
\author{O. Kaczmarek\\
        Fakult\"at f\"ur Physik, Universit\"at Bielefeld, D-33615 Bielefeld, Germany\\
        E-mail: \email{okacz@physik.uni-bielefeld.de}}  
  
\author{\speaker{Alexander Rothkopf}\\
        Institute for Theoretical Physics,  Heidelberg
  University, Philosophenweg 16, D-69120 Heidelberg, Germany\\
        E-mail: \email{rothkopf@thphys.uni-heidelberg.de}}      
        
\abstract{We present our latest results for the the complex valued static heavy-quark potential at finite temperature from lattice QCD. The real and imaginary part of the potential are obtained from the position and width of the lowest lying peak in the spectral function of the Wilson line correlator in Coulomb gauge. Spectral information is extracted from Euclidean time data using a novel Bayesian approach different from the Maximum Entropy Method. In order to extract both the real and imaginary part, we generated anisotropic quenched lattices $32^3\times N_\tau$ $(\beta=7.0,\xi=3.5)$ with $N_\tau=24,\ldots,96$, corresponding to $839{\rm MeV} \geq T\geq 210 {\rm MeV}$. For the case of a realistic QCD medium with light u,d and s quarks we use isotropic $48^3\times12$ ASQTAD lattices with $m_l=m_s/20$ provided by the HotQCD collaboration, which span $286 {\rm MeV} \geq T\geq 148{\rm MeV}$. We find a clean transition from a confining to a Debye screened real part and observe that its values lie close to the color singlet free energies in Coulomb gauge. The imaginary part, estimated on quenched lattices, is found to be of the same order of magnitude as in hard-thermal loop (HTL) perturbation theory.}

\FullConference{The 32nd International Symposium on Lattice Field Theory,\\
		23-28 June, 2014\\
		Columbia University New York, NY}

\begin{document}

\vspace{-0.3cm}
\section{Introduction}
\vspace{-0.3cm}

The experimental investigation of relativistic heavy-ion collisions has entered a precision era at RHIC and LHC. In particular the yields of the bound states of a heavy quark and anti-quark \cite{Brambilla:2010cs}, have been measured in the presence of a quark-gluon plasma in unprecedented detail (for $b\bar{b}$ e.g. see \cite{Chatrchyan:2012np,Abelev:2012rv}). The aim for theory is to provide a dynamical picture of heavy quarkonium evolution in a thermal environment, which in turn will allow us to deduce from the measured yields the properties of the surrounding medium. Lattice QCD plays an important role in this endeavor, as it is currently the only first principles method that can provide us with the necessary, non-perturbative information about the QCD medium at phenomenologically relevant temperatures just above the deconfinement transition.

Due to the separation of scales between the heavy quark mass and the typical energy density of the plasma created in current generation heavy-ion colliders, the real-time evolution of a heavy $Q\bar{Q}$ pair can be summarized in a Schroedinger equation with an effective interaction potential $V(r)$. With the help of effective field theories \cite{Brambilla:2004jw} this potential can be derived directly from the underlying field theory QCD. It was found to be related to the late real-time behavior of the thermal rectangular Wilson loop
\beq
V(r)=\lim_{t\to\infty} \frac{i\partial_t W(t,r)}{W(t,r)}\label{Eq:VRealTimeDef}, \quad W(t,r) = \left\langle {\rm exp}\Big[ - {ig}\int_{\square} dx_\mu A^\mu(x) \Big] \right\rangle .
\eeq
The evaluation of this expression in HTL perturbation theory \cite{Laine:2007qy} showed that the potential is complex valued. The real part exhibits Debye screening, while the imaginary part has been related to the scattering with (Landau damping) and absorption of (singlet-octet transition) gluons from the surrounding medium\footnote{A dynamical interpretation in the framework of open-quantum systems has been proposed in \cite{Akamatsu:2011se}\vspace{-0.6cm}}. For more than two decades the temperature dependence of the real part of this potential has been modeled \cite{Nadkarni:1986as} using either the color singlet free energies $F^{(1)}(r)$, the internal energies $U^{(1)}(r)$ or combinations thereof. Even though it has been shown that to leading order in resummed HTL perturbation theory ${\rm Re}[V]$ and $F^{(1)}$ agree, this relation already does not seem to hold at next to leading order \cite{Burnier:2009bk}. Hence a fully non-perturbative evaluation of Eq.\eqref{Eq:VRealTimeDef} is called for.

Over the past two years there has been steady progress, both conceptual and technical, towards a reliable evaluation of Eq.\eqref{Eq:VRealTimeDef} on the lattice. Since Monte Carlo simulations are performed in Euclidean time, the real-time Wilson loop is not directly accessible. The necessary information can nevertheless be obtained through the use of a spectral decomposition \cite{Rothkopf:2009pk,Rothkopf:2011db}
\begin{eqnarray}
 \nonumber W(\tau,r)=\int d\omega e^{-\omega \tau} \rho(\omega,r)\,
\leftrightarrow\, \int d\omega e^{-i\omega t} \rho(\omega,r)= W(t,r).
\end{eqnarray}
If combined with Eq.\eqref{Eq:VRealTimeDef} it connects the spectral function $\rho(\omega,r)$, which is accessible on the lattice in principle, to the potential
\begin{align}
\hspace{-0.2cm}V(r)=\lim_{t\to\infty}\int d\omega\, \omega e^{-i\omega t} \rho(\omega,r)/\int d\omega\, e^{-i\omega t} \rho(\omega,r). \label{Eq:PotSpec}
\end{align}
Two challenges need to be overcome in practice. First, spectral information needs to be inferred from a finite set of noisy Euclidean data points for $W(\tau_n,r),~n=1..N_\tau$. This is in general an ill-defined problem but can be given meaning through the use of Bayesian inference. In  this well established statistical approach, additional prior information is used to regularize an otherwise under-determined $\chi^2$ fit. For our study we use a novel Bayesian prescription for the reconstruction of $\rho(\omega,r)$ \cite{Burnier:2013nla} that has been shown in mock data tests to surpass the usual Maximum Entropy Method \cite{Asakawa:2000tr} in both accuracy and precision.

\begin{figure}[t]
\centering \vspace{-0.6cm}
 \includegraphics[scale=0.49]{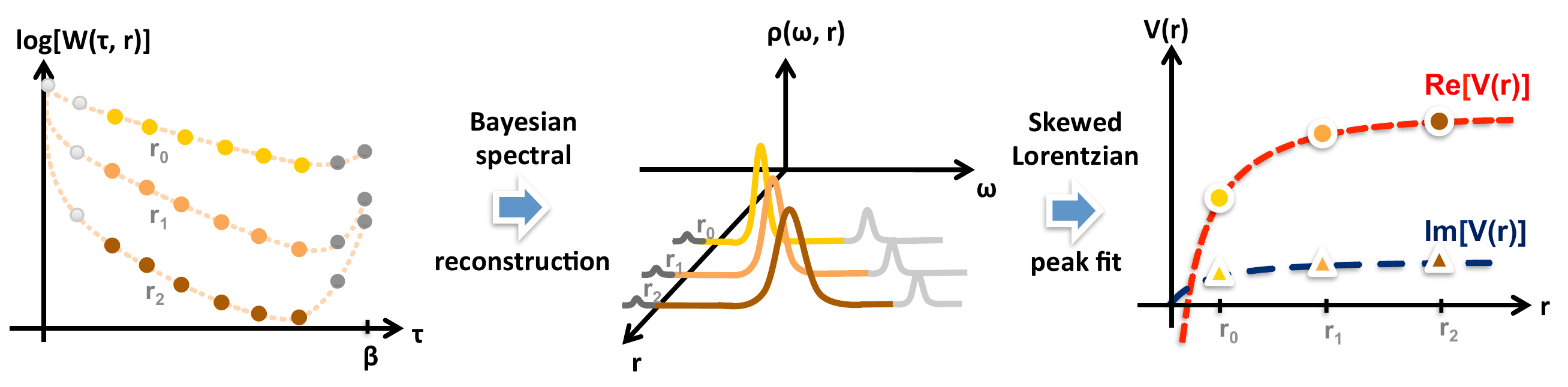} \vspace{-0.3cm}
 \caption{From Euclidean lattice QCD correlators to the complex valued static heavy quark potential.}\label{Fig1} \vspace{-0.3cm}
\end{figure}

The second challenge is related to carrying out the late time limit in Eq.\eqref{Eq:PotSpec}. In \cite{Rothkopf:2011db} it was noted that the lowest lying peak in $\rho(\omega,r)$ will ultimately dominate the dynamics, so that one can directly fit its shape and carry out the Fourier transform of Eq.\eqref{Eq:PotSpec} analytically. The exact functional form to be used in such a fit was however only understood later on in \cite{Burnier:2012az} and found to be a skewed Lorentzian
\begin{align}
\rho\notag\propto\frac{|{\rm Im} V(r)|{\rm cos}[{\rm Re}{\sigma_\infty}(r)]-({\rm Re}V(r)-\omega){\rm sin}[{\rm Re} {\sigma_\infty}(r)]}{ {\rm Im} V(r)^2+ ({\rm Re} V(r)-\omega)^2}+{c_0}(r)+{c_1}(r)({\rm Re} V(r)-\omega)+\ldots\,.
\end{align}
The position of this peak encodes the real part, the width on the other hand determines the imaginary part of the potential. The feasibility of this procedure, sketched in  Fig.\ref{Fig1}, has been tested using HTL perturbation theory in \cite{Burnier:2013fca}. Indications were found that for the extraction of the potential alone we do not need to rely on the rather noisy Wilson loop, but can use the the Wilson line correlators $W_{||}(\tau,r)$ in Coulomb gauge instead. The reason for the generally better signal to noise ratio is the absence of cusp divergences in the latter (see \cite{Berwein:2012mw} and references therein).

\vspace{-0.1cm}
\section{Quenched lattice QCD}
\vspace{-0.1cm}

\begin{table}[b!]
\vspace{-0.3cm}

\begin{tabularx}{15cm}{ | c | X | X | X | X | X | X | X | X | X | }
\hline
	SU(3):$N_\tau$ & 24 & 32 & 40 & 48 & 56 & 64 & 72 & 80 & 96 \\ \hline
	$T$[MeV] & 839 & 629 & 503 & 419 & 360 & 315 & 280 & 252 & 210 \\ \hline
	$N_{\rm meas}$ & 3270 & 2030 & 1940 & 1110 & 1410 &1520 & 860 & 1190 & 1800 \\ \hline
\end{tabularx}
\caption{Quenched SU(3) on $32^3\times N_\tau$ anisotropic $\xi_b=3.5$ lattices with $a_s=0.039$fm and  $T_c\approx271$MeV.}\label{Tab:SU3LatParm}
\vspace{-0.6cm}
\end{table}

We begin the presentation of our results with the temperature dependence of the inter-quark potential on quenched lattices. In addition to \cite{Burnier:2014ssa} we also show some explicit consistency checks. In the absence of dynamical fermions, the number of temporal lattice points can easily surpass $N_\tau=24$, which allows for a reliable determination of the real part and a robust order of magnitude estimate for the imaginary part. Using the naive anisotropic Wilson action with $\beta=7$ and $\xi=3.5$ \cite{Asakawa:2003re} we generated $32^3\times N_\tau$ configurations, changing temperature $839{\rm MeV} (3.11T_c) \geq T\geq 210 {\rm MeV} (0.78T_c)$ by varying the temporal extend between $24 \geq N_\tau \geq 96$ (see Tab.\ref{Tab:SU3LatParm}). To measure the Wilson line correlators we iteratively fix to Coulomb gauge using Fourier acceleration, and evaluate the observable along each spatial axis on the square- and cubic diagonals. The spatial distances are corrected for lattice artifacts from a comparison between free propagators on the lattice and in the continuum \cite{Necco:2001xg}.

\begin{figure}[t]
\centering \vspace{-0.4cm}
 \includegraphics[scale=0.43]{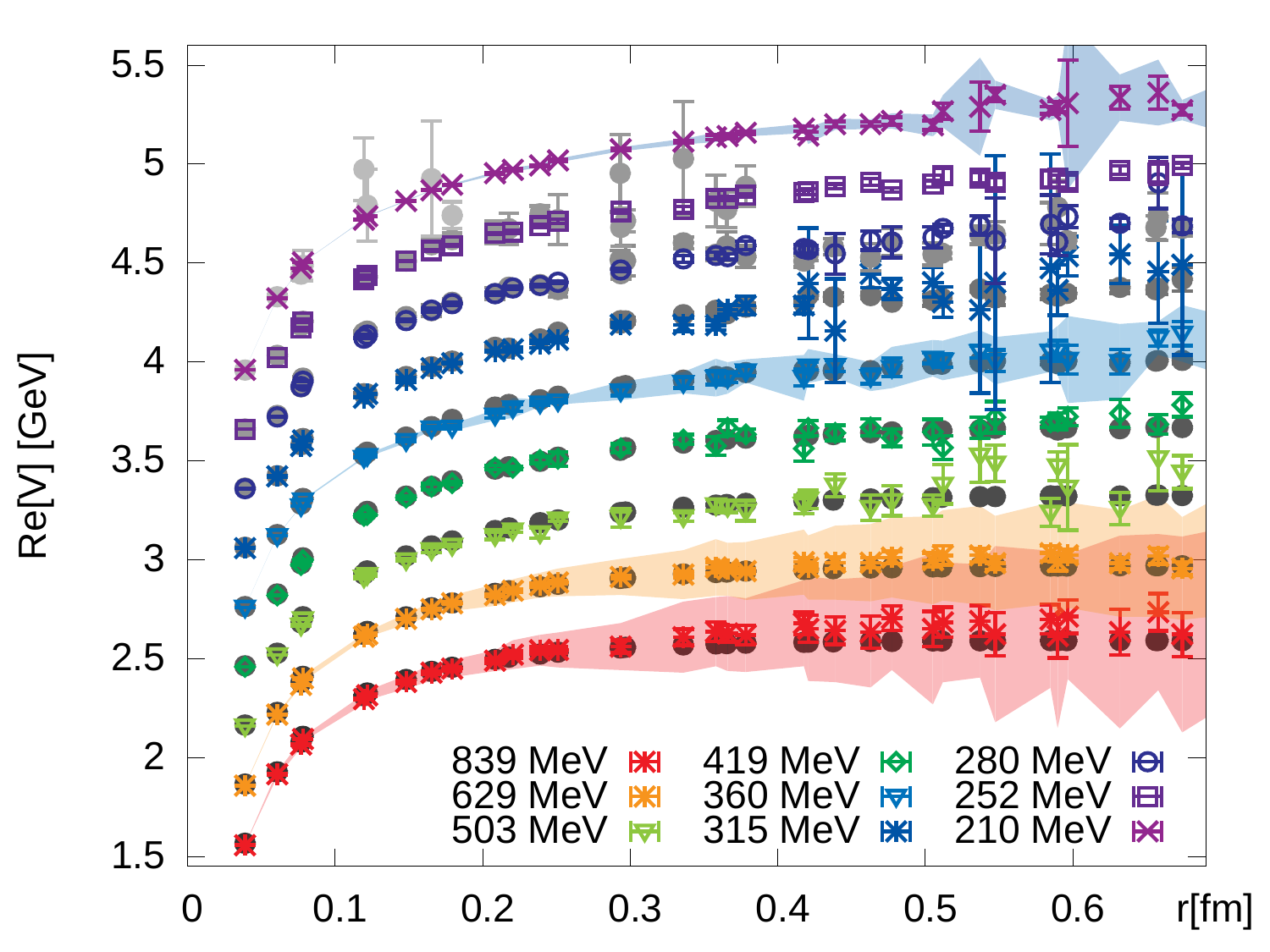} \includegraphics[scale=0.43]{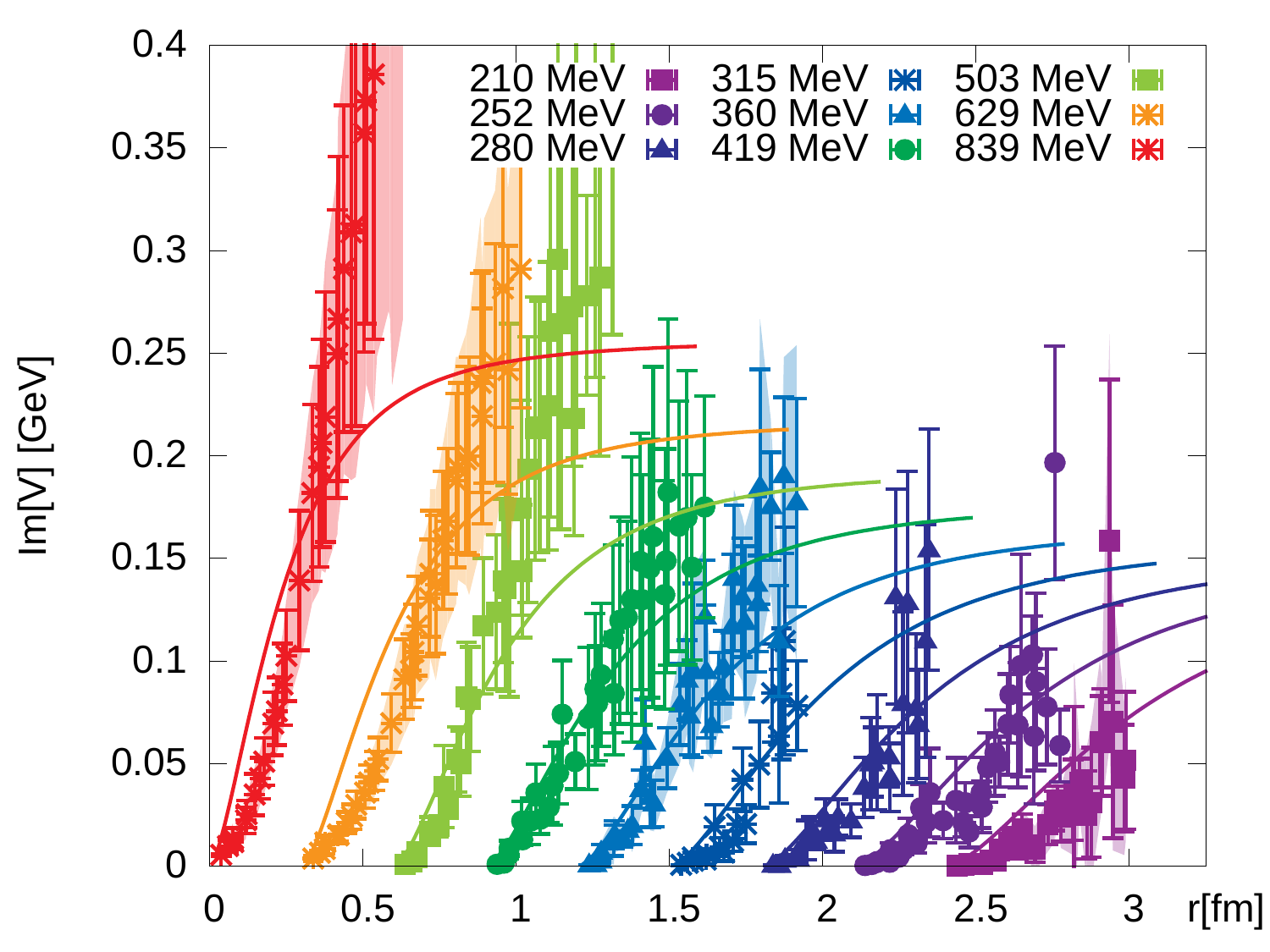} \vspace{-0.3cm}
 \caption{Gluonic medium: (left) Comparison between the real part of the static inter-quark potential (open symbols) and the color singlet free energies (gray circles). The values are shifted for better readability. Error bars are obtained from Jackknife variance, error bands from additional systematics as described in the text. (right) Horizontally shifted values of ${\rm Im}[V]$ (symbols) compared to leading order hard-thermal loop perturbation theory (solid lines).}\label{Fig2} 
\end{figure}

The spectral functions underlying the potential are extracted from all imaginary time datapoints except those at $\tau=0, \beta$. By this we avoid possible overlap divergences \cite{Berwein:2012mw}. Our frequency range is chosen to lie between $\omega^{\rm num}\in[-168,185]\times N_\tau/24$ GeV using $N_\omega=4000$ points, among which $N_{\rm hr}=550$ are used to resolve the lowest lying peak. For a neutral reconstruction we deploy a flat default model $m(\omega)={\rm const.}$ and use $512$bit precision arithmetic in a LBFGS minimizer to find the unique Bayesian spectral function related to the measured data. Our stopping criterion is a step size of $\Delta=10^{-60}$. The subsequent fit of a skewed Lorentzian shape around the full width at half maximum of the lowest lying peak yields the values for Re[$V$] and Im[$V$] shown in Fig.\ref{Fig2}. The error bars are obtained from the Jackknife variance between ten reconstructions, each excluding  a consecutive block of 10\% of the measurements $N_{\rm meas}$. 

\begin{figure}[t]
\centering \vspace{-0.3cm}
 \includegraphics[scale=0.43]{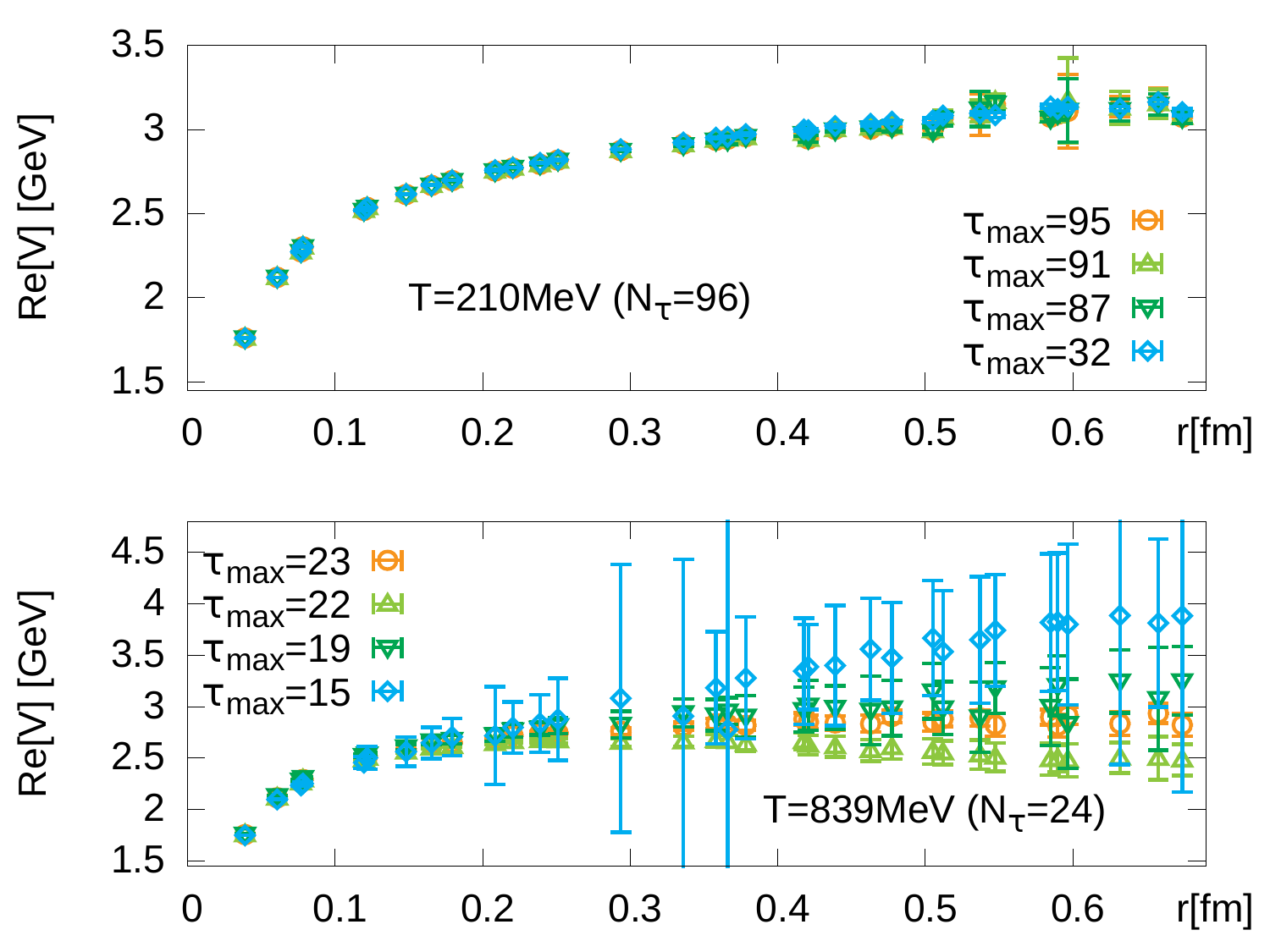} \includegraphics[scale=0.43]{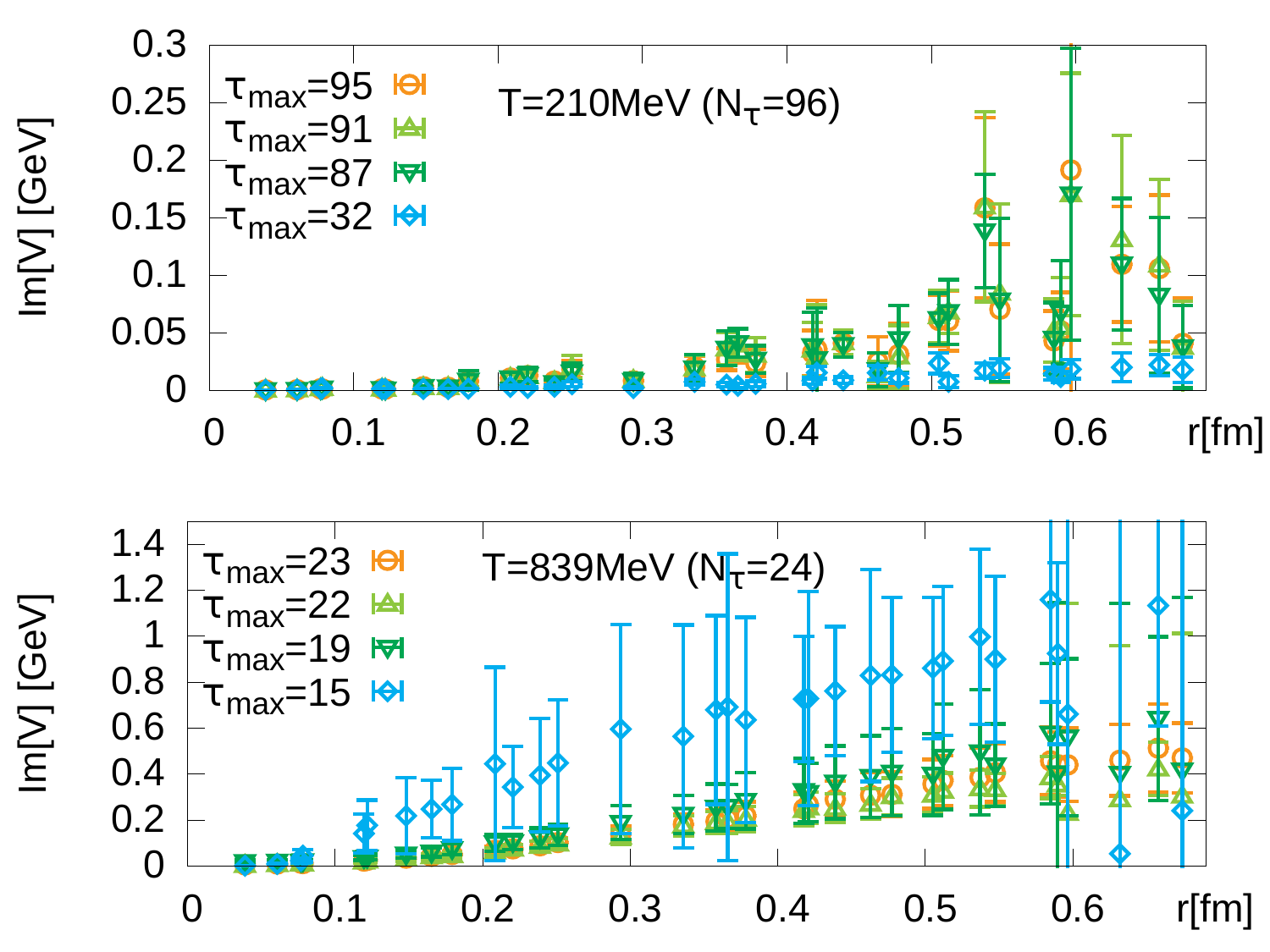}\vspace{-0.15cm}
 \caption{Dependence of Re[V] (left) and Im[V] (right) on the removal of datapoints close to $\tau=\beta$. Even with $\tau_{\rm max}=32$ for $N_\tau=96$ we do not find a downward trend in Re[V] that would mimic Debye screening at higher T. On the other hand beyond $\tau_{\rm max}\leq15$ the precision of the reconstruction degrades significantly.}\label{Fig3} \vspace{-0.3cm}
\end{figure}

As expected, the confining linear rise of the real part in the hadronic phase goes over smoothly into a Debye screened form above the deconfinement temperature. When comparing Re[V] (open colored symbols) to the color singlet free energies $F^{(1)}(r)=-T{\rm log}[W_{||}(r,\tau=\beta)]$ (gray filled circles) we find that their values lie close to each other at all temperatures investigated (perturbation theory indicates \cite{Burnier:2009bk} that at intermediate distances Re[V] will lie slightly below $F^{(1)}(r)$). The fact that the spectral reconstruction takes into account several points along $\tau$, while the free energies use only $\tau=\beta$, explains why our values for Re[V] at $T\simeq T_c$ show a much smaller variance. At the same time Im[V] is found to be of the same order of magnitude as the  estimates from leading order hard-thermal loop resummed perturbation theory. This result for Im[V] is consistent with a recent modeling study which deploys HTL spectral functions to fit the Wilson line correlators \cite{Bazavov:2014kva}.

The results of a spectral reconstruction can suffer from several systematic factors. To quantify these uncertainties we give in Fig.\ref{Fig2} an error band that is obtained from the maximum variation among reducing the number of datapoints along $\tau$ by four or eight, changing the default model dependence normalization ($\times 10$, $\times 0.1$) or functional form ($m\propto{\rm const},\omega^{-2},\omega^2$) as well as removing 10\%, 20\% and 30\% of the statistics of the input data. The number of available datapoints is particularly important, which is why we show its influence explicitly in in Fig.\ref{Fig3} for the lowest ($T=210$MeV, $N_\tau=96$) and highest ($T=839$MeV, $N_\tau=24$) temperature. Note that taking into account only $\tau_{\rm max}=32$ at $N_\tau=96$ does not change Re[V] beyond statistical errors. This is why we attribute the change in the functional form of Re[V] towards Debye screening at higher temperatures indeed to the effects of a thermal medium. At the same time it appears that Im[V] shrinks, which is not surprising, since a finite spectral width mostly dominates the correlator at late $\tau$. If such values are discarded it enters at intermediate $\tau$ only exponentially suppressed. On the other hand, Im[V] can be extracted even at relatively small $N_\tau=24$ if the late $\tau$ values are retained. Reducing further to $\tau_{\rm max}=16$ starts to significantly increase the variance of the reconstruction, the correct result however still lies within the Jackknife error bars.

\vspace{-0.2cm}
\section{Dynamical lattice QCD}
\vspace{-0.2cm}

\begin{table}[b!]
\vspace{-0.3cm}
\begin{tabularx}{15cm}{ | c | X | X | X | X | X | X | X| }
\hline
	QGP: $\beta$ \hspace{0.8cm}& 6.8 & 6.9 & 7 & 7.125 & 7.25 & 7.3 & 7.48 \\ \hline
	$T$[MeV] & 148 & 164 & 182 & 205 & 232 & 243 & 286 \\ \hline
	a [fm] & 0.111 & 0.1 & 0.09 & 0.08 & 0.071 & 0.068 & 0.057\\ \hline
	$N_{\rm meas}$ & 1295 & 1340 & 1015 & 840 & 1220 & 1150 & 1130 \\ \hline
\end{tabularx}
\caption{The isotropic HotQCD $48^3\times12$ lattices with ASQTAD action ($m_l=m_s/20,T_c\approx174$MeV).}\label{Tab:DynLatParm}
\vspace{-0.6cm}
\end{table}

\begin{figure}[t]
\centering \vspace{-0.75cm}
 \includegraphics[scale=0.43]{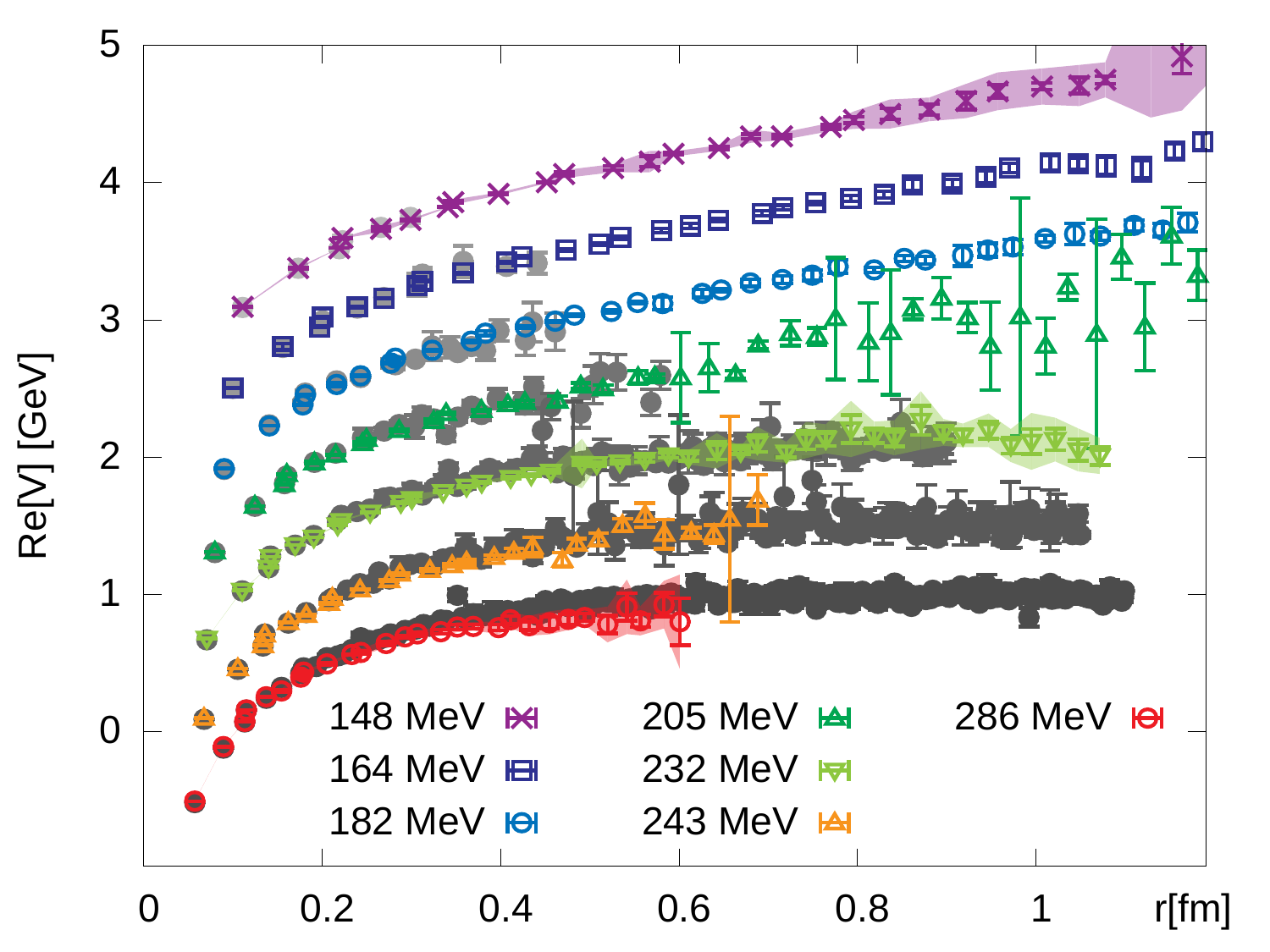}\includegraphics[scale=0.43]{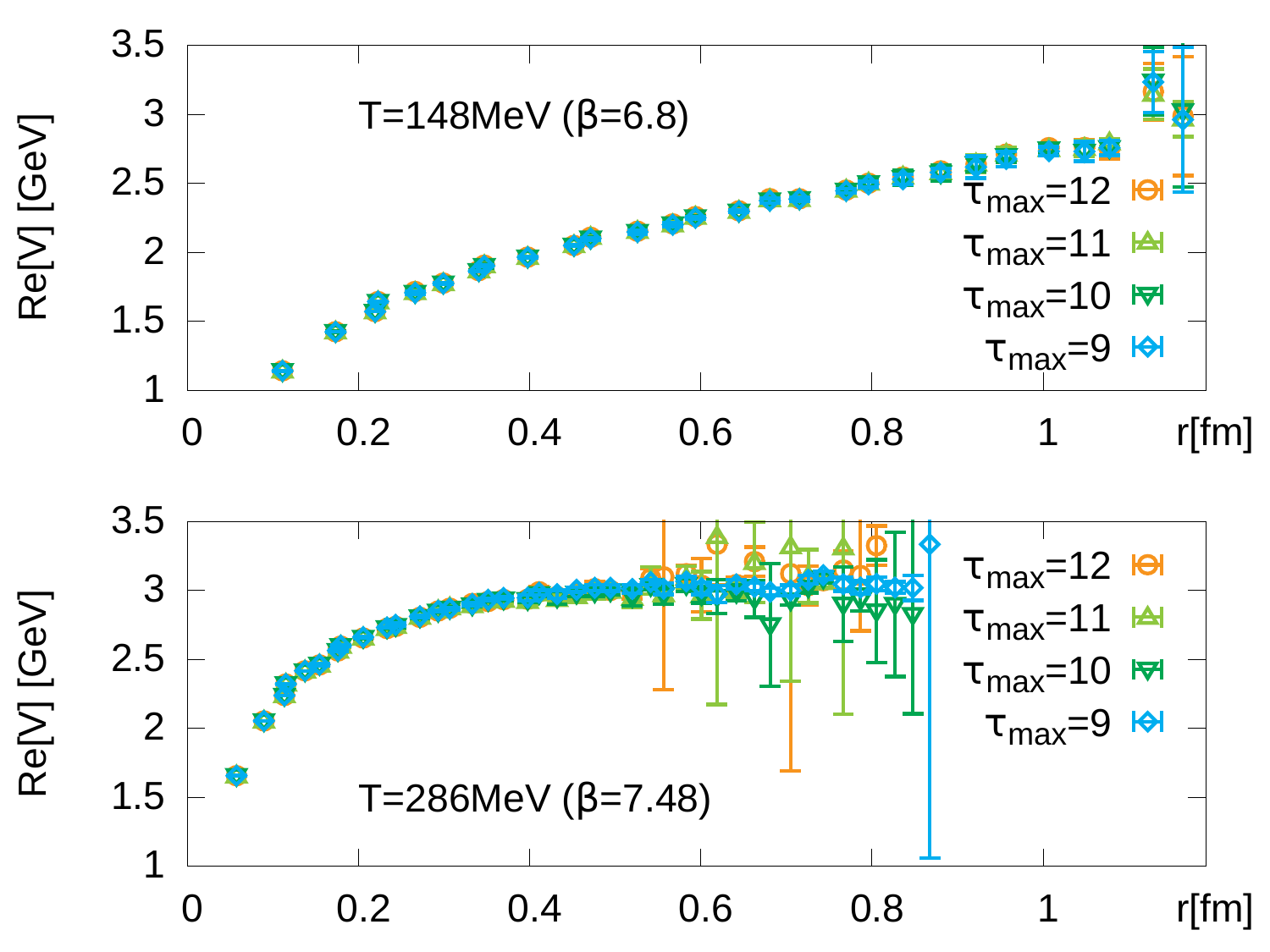}
 \caption{Full QCD: (left) Comparison between the real part of the static inter-quark potential (open symbols) and the color singlet free energies (gray circles). The values are shifted for better readability. (right) Results for Re[V] for different number of datapoints at the lowest ($\beta=6.8$) and highest ($\beta=7.48$) temperatures. Since the datapoints close to $\tau=\beta$ have a relatively low signal-to noise ratio their removal reduces the Jackknife variance of the result at the highest T.}\label{Fig4} \vspace{-0.5cm}
\end{figure}

For phenomenological applications the behavior of the potential in a realistic QCD medium needs to be understood. We proceed towards this goal by carrying out the extraction procedure laid out above on dynamical lattices generated by the HotQCD collaboration \cite{Bazavov:2011nk}, which contain light u, d and s quarks. The temperature on these isotropic $48^3\times12$ lattices with the $N_f=2+1$ ASQTAD action ($m_l=m_s/20$) is changed between $286 {\rm MeV} (1.64T_c) \geq T\geq 148{\rm MeV} (0.85T_c)$ through varying the lattice spacing, so that at each T the same number of datapoints is available (see Tab.\ref{Tab:DynLatParm}). Unfortunately $N_\tau=12$ is not large enough to obtain reliable values for the spectral width, i.e. Im[V], so that we show only the real-part of the potential in Fig.\ref{Fig4}. The spectral reconstruction is performed using $\beta^{\rm num}=20$ and a numerical frequency interval between $\omega\in[-11,12]$, which is divided in $N_\omega=4600$ steps. $N_{\rm hr}=1000$ points are used to resolve the lowest lying peak in detail.

We find that the spectral based determination of Re[V] is very robust at $T\simeq T_c$ and allows us to go up to distances $r\simeq 1.2$fm. We do not see any indication of string breaking yet, probably since the pion masses on these lattices $M_\pi\approx 300$MeV are still larger than the physical value. The presence of dynamical fermions significantly changes the location of the phase transition compared to pure SU(3) theory, so that here already at $T=286$MeV Debye screening is well pronounced. However, just as in the quenched case, the values of the color singlet free energies (gray filled circles) lie close to Re[V].

The error bars in Fig.\ref{Fig4} are obtained from the Jackknife variance between ten individual reconstructions, each with a different set of 10\% of the statistics removed. The error bands are based on the maximum variance between changing the number of datapoints in the reconstruction by one and two, changing the default model dependence normalization ($\times 10$, $\times 0.1$) or functional form ($m\propto{\rm const},\omega^{-2},\omega^2$) as well as removing 10\%, 20\% and 30\% of the statistics of the input data.

Since we only have at our disposal a small number of datapoints, we show explicitly on the right of Fig.\ref{Fig4} how the results changes if additional points are removed. Due to the difference in lattice spacing the removal of one temporal step at $\beta=6.8$ amounts to a larger cut in physical extend than for $\beta=7.48$. We find that for the statistics and the lattice spacing available the effects are small enough that it is possible to attribute the changes between the lowest temperature (confining linear rise) and the highest temperature (Debye screened) indeed to medium effects.

\vspace{-0.5cm}
\section{Conclusion}
\vspace{-0.25cm}

The spectral function based determination of the real-time potential from lattice QCD has matured due to recent conceptual and technical progress. A crucial ingredient is a novel Bayesian spectral reconstruction prescription that allows to faithfully reconstruct the skewed Lorentzian shapes encoded in the spectra of Wilson loop and Wilson line correlation functions from Euclidean time simulations. We have determined the values of both Re[V] and Im[V] in a purely gluonic medium and the real-part in a realistic QCD medium with light u, d and s quarks. The main finding is that Re[V] lies close to the color singlet free energies and thus shows a smooth transition from the linearly rising confining behavior towards Debye screening. The imaginary part is found to be of the same order of magnitude as in hard-thermal loop resummed perturbation theory. The authors thank H.~B. Meyer, M.~P. Lombardo, P. Petreczky and J.-I. Skullerud for fruitful discussions. YB is supported by SNF grant PZ00P2-142524.


\vspace{-0.4cm}
\begin{thebibliography}{99}
\vspace{-0.15cm}

\bibitem{Brambilla:2010cs} 
  N.~Brambilla {\it et al.},
  ``Heavy quarkonium: progress, puzzles, and opportunities,''
  Eur.\ Phys.\ J.\ C {\bf 71}, 1534 (2011).
  
\bibitem{Chatrchyan:2012np}
  S.~Chatrchyan {\it  et al.}  [CMS Collaboration],
  ``Suppression of non-prompt $J/\psi$, prompt $J/\psi$, and Y(1S) in PbPb collisions at $\sqrt{s_{NN}}=2.76$ TeV,''
  JHEP {\bf 1205} (2012) 063

\bibitem{Abelev:2012rv}  
  B.~Abelev {\it et al.}  [ALICE Collaboration],
  ``$J/\psi$ suppression at forward rapidity in Pb-Pb collisions at $\sqrt{s_{NN}}=2.76$ TeV,''
  Phys.\ Rev.\ Lett.\  {\bf 109} (2012) 072301.
  
\bibitem{Brambilla:2004jw} 
  N.~Brambilla, A.~Pineda, J.~Soto and A.~Vairo,
  ``Effective field theories for heavy quarkonium,''
  Rev.\ Mod.\ Phys.\  {\bf 77}, 1423 (2005).

\bibitem{Laine:2007qy} 
  M.~Laine, O.~Philipsen and M.~Tassler,
  ``Thermal imaginary part of a real-time static potential from classical lattice gauge theory simulations,''
  JHEP {\bf 0709}, 066 (2007);
  A.~Beraudo, J.~-P.~Blaizot, C.~Ratti,
  ``Real and imaginary-time Q anti-Q correlators in a thermal medium,''
  Nucl.\ Phys.\ A {\bf 806}, 312 (2008);
  N.~Brambilla, J.~Ghiglieri, A.~Vairo, P.~Petreczky,
  ``Static quark-antiquark pairs at finite temperature,''
  Phys.\ Rev.\ D {\bf 78}, 014017 (2008).
  
\bibitem{Akamatsu:2011se} 
  Y.~Akamatsu and A.~Rothkopf,
  ``Stochastic potential and quantum decoherence of heavy quarkonium in the quark-gluon plasma,''
  Phys.\ Rev.\ D {\bf 85}, 105011 (2012);

\bibitem{Nadkarni:1986as} 
  S.~Nadkarni,
  ``Nonabelian Debye Screening. 2. The Singlet Potential,''
  Phys.\ Rev.\ D {\bf 34}, 3904 (1986);
  C.~Y.~Wong,
  ``Heavy quarkonia in quark-gluon plasma,''
  Phys.\ Rev.\ C {\bf 72}, 034906 (2005);
  O.~Kaczmarek,
  ``Screening at finite temperature and density,''
  PoS CPOD {\bf 07}, 043 (2007);
  H.~Satz,
  ``Heavy Quark Interactions and Quarkonium Binding,''
  J.\ Phys.\ G {\bf 36}, 064011 (2009).
  
\bibitem{Burnier:2009bk} 
  Y.~Burnier, M.~Laine and M.~Vepsalainen,
  ``Dimensionally regularized Polyakov loop correlators in hot QCD,''
  JHEP {\bf 1001}, 054 (2010).
 
\bibitem{Rothkopf:2009pk} 
  A.~Rothkopf, T.~Hatsuda and S.~Sasaki,
  ``Proper heavy-quark potential from a spectral decomposition of the thermal Wilson loop,''
  PoS LAT {\bf 2009}, 162 (2009);
  
\bibitem{Rothkopf:2011db} 
  A.~Rothkopf, T.~Hatsuda and S.~Sasaki,
  ``Complex Heavy-Quark Potential at Finite Temperature from Lattice QCD,''
  Phys.\ Rev.\ Lett.\  {\bf 108}, 162001 (2012)
  
\bibitem{Burnier:2013nla}
  Y.~Burnier and A.~Rothkopf,
  ``Bayesian Approach to Spectral Function Reconstruction for Euclidean Quantum Field Theories,''
  Phys.\ Rev.\ Lett.\  {\bf 111} (2013) 18,  182003;
  PoS LATTICE {\bf 2013} (2013) 490.

\bibitem{Asakawa:2000tr}
  M.~Asakawa, T.~Hatsuda and Y.~Nakahara,
  ``Maximum entropy analysis of the spectral functions in lattice QCD,''
  Prog.\ Part.\ Nucl.\ Phys.\  {\bf 46} (2001) 459;
  A.~Rothkopf,
  ``Improved Maximum Entropy Analysis with an Extended Search Space,''
  J.\ Comput.\ Phys.\  {\bf 238} (2013) 106.
  
\bibitem{Burnier:2012az}
  Y.~Burnier and A.~Rothkopf,
  ``Disentangling the timescales behind the non-perturbative heavy quark potential,''
  Phys.\ Rev.\ D {\bf 86} (2012) 051503.
 
\bibitem{Burnier:2013fca}
  Y.~Burnier and A.~Rothkopf,
  ``A hard thermal loop benchmark for the extraction of the nonperturbative $Q\bar{Q}$ potential,''
  Phys.\ Rev.\ D {\bf 87} (2013) 11,  114019;
  ``Benchmarking the Bayesian reconstruction of the non-perturbative heavy $Q\bar{Q}$ potential,''
  PoS LATTICE {\bf 2013} (2013) 491.
 
\bibitem{Berwein:2012mw}
  M.~Berwein, N.~Brambilla, J.~Ghiglieri and A.~Vairo,
  ``Renormalization of the cyclic Wilson loop,''
  JHEP {\bf 1303} (2013) 069.

\bibitem{Burnier:2014ssa} 
  Y.~Burnier, O.~Kaczmarek and A.~Rothkopf,
  ``Static quark-antiquark potential in the quark-gluon plasma from lattice QCD,''
  arXiv:1410.2546 [hep-lat].
   
\bibitem{Asakawa:2003re} 
  M.~Asakawa and T.~Hatsuda,
  ``J / psi and eta(c) in the deconfined plasma from lattice QCD,''
  Phys.\ Rev.\ Lett.\  {\bf 92}, 012001 (2004).

\bibitem{Necco:2001xg}
  S.~Necco and R.~Sommer,
  ``The N(f) = 0 heavy quark potential from short to intermediate distances,''
  Nucl.\ Phys.\ B {\bf 622} (2002) 328.
  
\bibitem{Bazavov:2014kva}
  A.~Bazavov, Y.~Burnier and P.~Petreczky,
  ``Lattice calculation of the heavy quark potential at non-zero temperature,''
  arXiv:1404.4267 [hep-lat].
  
\bibitem{Bazavov:2011nk} 
  A.~Bazavov {\it et al.},
  ``The chiral and deconfinement aspects of the QCD transition,''
  Phys.\ Rev.\ D {\bf 85}, 054503 (2012)
  
\end{thebibliography}
\end{document}